\documentclass[aps,floatfix, twocolumn]{revtex4} 
\usepackage{graphicx}
\usepackage{dcolumn}
\usepackage{bm}
\usepackage{longtable}
 \usepackage{amsmath}
 \usepackage{amssymb}
 \usepackage{eurosym}
\usepackage{float}
\begin{document}

\title{Disentangling genetic and environmental risk factors for individual diseases from multiplex comorbidity networks}

\author{Peter Klimek$^1$, Silke Aichberger$^1$, Stefan Thurner$^{1,2,3,}$}
\email{stefan.thurner@meduniwien.ac.at}

\affiliation{$^1$Section for Science of Complex Systems; Medical University of Vienna; 
Spitalgasse 23; A-1090; Austria\\ 
$^2$Santa Fe Institute; 1399 Hyde Park Road; Santa Fe; NM 87501; USA\\
$^3$IIASA, Schlossplatz 1, A 2361 Laxenburg; Austria}

\begin{abstract} 
Most disorders are caused by a combination of multiple genetic and/or environmental factors.
If two diseases are caused by the same molecular mechanism, they tend to co-occur in patients.
Here we provide a quantitative method to disentangle how much genetic or environmental risk factors contribute to the pathogenesis of 358 individual diseases, respectively.
We pool data on genetic, pathway-based, and toxicogenomic disease-causing mechanisms with disease co-occurrence data obtained from almost two million patients.
From this data we construct a multilayer network where nodes represent disorders that are connected by links that either represent phenotypic comorbidity of the patients or the involvement of a certain molecular mechanism.
From the similarity of phenotypic and mechanism-based networks for each disorder we derive  measure that allows us to quantify the relative importance of various molecular mechanisms for a given disease.
We find that most diseases are dominated by genetic risk factors, while environmental influences prevail for disorders such as depressions, cancers, or dermatitis.
Almost never we find that more than one type of mechanisms is involved in the pathogenesis of diseases. 
\end{abstract}
 
\maketitle


\section{Introduction}

\begin{figure}[tbp]
\begin{center}
 \includegraphics[width = 0.35\textwidth, keepaspectratio = true]{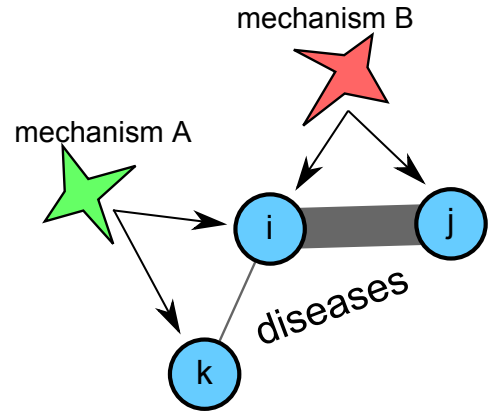}
\end{center}
 \caption{Consider three diseases $i$, $j$, $k$ (blue circles) and assume that disease $i$ co-occurs very frequently with $j$ (thick line) but only in rare cases with $k$ (thin line). Further, assume that there are two different disease-causing mechanisms for $i$, $A$ and $B$, where mechanism $A$ ($B$) is also known to be involved in disease $k$ ($j$). Since $i$ is very often observed together with $j$, but not with $k$, mechanism $B$ explains the disease phenotype $i$ much better than $A$.}
 \label{fig1}
\end{figure}

Multifactorial diseases are disorders that involve multiple disease-causing mechanisms, such as genes acting in concert with environmental factors.
The represent one of the most significant challenges that medical research faces today \cite{Lim12}.
Disease-causing mechanisms may be (and typically are) involved in more than one disorder \cite{Barabasi11}.
If two diseases are related to the same mechanism (say, a single point mutation, SNP, or an altered metabolic pathway), they have a tendency to co-occur in the same patients \cite{Rzhetsky07, Lee08}.
Here we develop a novel network-medicine approach to quantify the relative contributions of genetic and environmental risk factors for diseases.
The central idea of the approach is illustrated in figure \ref{fig1}.
We consider three diseases $i$, $j$, $k$ (circles) and assume that diseases $i$ and $j$ co-occur very frequently in patients (thick line), whereas diseases $i$ and $k$ rarely coincide within patients (thin line).
Assume further that $i$ can arise through two different disease-causing mechanisms, $A$ and $B$, where mechanism $A$ is also responsible for (or involved in) disease $k$ and mechanism $B$ for disease $j$.
Obviously, mechanism $B$ explains the observed disease phenotype $i$ (the frequent co-occurrence with disease $j$) much better than mechanism $A$ and is therefore a more probable causes for disease $i$.
Using this idea we are able to identify the most likely causes and are able to disentangle genetic and environmental disease-causing mechanisms for 358 different disease phenotypes.
 
We consider the three most important classes of disease-causing mechanisms.
(i) {\it Genetic} mechanisms relate a disease to a specific defect or alteration in the genome.
If one such defect is related to two or more pathologies, then those diseases share a genetic comorbidity.
For example, it was shown that the phenotypic comorbidity between schizophrenia and Parkinson's disease is almost entirely accounted for by SNPs in loci near NT5C2 and HLA-DRA \cite{Nalls14}. 
(ii) {\it Pathway-based} mechanisms are given by a defective pathway (e.g. metabolic or signal transduction pathway) that is involved in the etiology of the disease.
Pathway-based comorbidities indicate that two diseases are related to different defects in the same pathway.
For instance, it is known that the Pi3K/AKT pathway up-regulates anti-inflammatory cytokines and inhibits proinflammatory cytokines such as IL-1b, IL-6, TNF-$\alpha$, and IFN-$\gamma$ that show increased levels in patients with major depressive disorder \cite{Kitagishi12}.
Also, inactivation of the Pi3K/AKT pathway through the suppression of insulin receptor substrates (IRS) may act as the underlying mechanism for the metabolic syndrome (i.e. the frequent concurrence of metabolic disorders such as hypertension, obesity, or diabetes) \cite{Guo14}. 
Indeed, depression has been identified as an important comorbidity of the metabolic syndrome in various cross-sectional surveys \cite{Dunbar08, Klimek15}.
Finally, (iii) {\it toxicogenomic} mechanisms characterize diseases caused by exposure to chemical substances that change the activity of certain genes.
Two diseases share a toxicogenomic comorbidity if they are related to different genes that interact with the same toxic substance.
For example, the immunosuppressive chemical methoxychlor is used as pesticide and can cause atopic dermatitis, possibly by expressing IL-13 in the skin \cite{Zhu11}.
Methoxychlor also promotes the epigenetic transgenerational inheritance of kidney disease.
Upon prenatal exposure to methoxychlor during fetal gonadal development, offspring show increased incidence of adult-onset kidney disease that was related to differentially DNA methylated regions \cite{Manikkam14}.
Atopic dermatitis is indeed associated with the nephritic syndrome \cite{Darlenski14}.
There may be cases where the same diseases share genetic and environmental risk factors.
Such cases we regard as {\it genetic} comorbidities, because the genetic link represents a direct mechanism that explains a corresponding phenotypic comorbidity without the need for additional, environmental influences.

The construction and analysis of networks of diseases that are connected by different comorbidity relations has recently lead to substantial progress in our understanding of the etiologies of various diseases \cite{Barabasi11, Pawson08, Zanzoni09}.
For instance, gene-disease associations collected in the Online Mendelian Inheritance in Man (OMIM) database \cite{OMIM} can be used to construct a network where diseases are linked if they are related to the same mutations in one or several genes \cite{Goh07}.
This network allowed for the identification of clusters of diseases, such as cancers, which are held together by a small number of genes \cite{Feldman08}.
Another approach is to connect diseases if they are both associated with enzymes that catalyze reactions in the same pathway \cite{Lee08}.
Protein-protein interaction data can be integrated with toxicogenomics data to construct a network where two diseases are linked if they are both caused by exposure to the same chemical, which has led to the successful identification of novel chemical-protein associations \cite{Audouze10}.
It has recently been shown that diseases that are comorbid in the population tend to be related with clusters of proteins that are close to each other in the human protein-protein interaction network \cite{Menche15}.
Different types of genomic, metabolomic, and proteomic disease-disease relations have also been combined to form an ``integrated disease network'' \cite{Sun14a, Sun14b}.
In phenotypic comorbidity networks, nodes correspond to disease phenotypes that are linked if the two diseases tend to co-occur in the same patients \cite{Hidalgo09}.
Chronic, multifactorial disorders often assume the role of hubs in such networks (i.e. nodes that are strongly connected with a large number of other diseases) \cite{Chmiel14}.

\begin{figure*}[tbp]
\begin{center}
 \includegraphics[width = 0.9\textwidth, keepaspectratio = true]{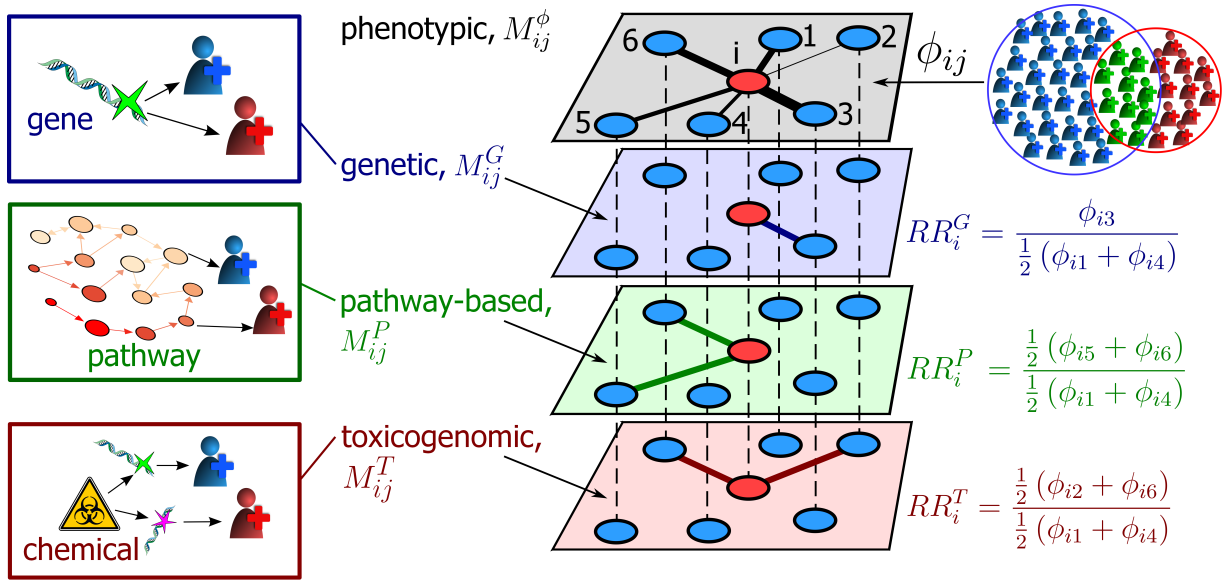}
\end{center}
 \caption{Illustration of the HDMN for a disease $i$. In the HDMN, nodes correspond to disease phenotypes that are connected by four different types of links which can be visualized as network layers. The first layer, $M_{ij}^{\phi}$, encodes phenotypic comorbidity relations. The link-weights in this layer are given by the comorbidity strengths $\phi_{ij}$ that measure how often two diseases $i$ and $j$ co-occur within the same patients, i.e.  the numbers of patients with either disease $i$ (red individuals) or $j$ (blue) are compared to the numbers of patients with both diseases (green). The second layer, $M_{ij}^G$, contains genetic comorbidities (blue links) where two different phenotypes (illustrated as blue and red individuals) are related to the same genetic defect or alteration. The third type of links are pathway-based comorbidities (green links), layer $M_{ij}^P$. Here, two different alterations occur in a pathway that is involved in two or more different diseases. Finally, the fourth layer, $M_{ij}^T$, is given by toxicogenomic comorbidities (red links), where a chemical substance is known to trigger different disease-causing mechanisms. Disorder $i$ is shown as a red node in the HDMN, together with other phenotypes (blue nodes) that are in $i$'s neighborhood in at least one of the layers. 
The relative comorbidity risks $RR_i^{\alpha}$ measure to which extent shared disease-causing mechanisms between two diseases lead to their phenotypic comorbidity.
$RR_i^{\alpha}$ is the average comorbidity strength of all neighbors of $i$ in layer $\alpha$, normalized to the average comorbidity strength over all phenotypes that share no disease-causing mechanism of any type with $i$.}
 \label{fig2}
\end{figure*}

Here we construct a generalized network that combines phenotypic comorbidity networks with those given by different types of shared disease-causing mechanisms (genes, pathways, or exposure to chemicals), the {\it human disease multiplex network} (HDMN) (see figure \ref{fig2}).
Multiplex networks are given by a set of nodes connected by multiple sets of links \cite{Boccaletti14, Kivela14}.
One set of links in the HDMN corresponds to phenotypic comorbidity relations, whereas the other sets of links represent different classes of genetic or environmental mechanisms.
We quantify how {\it similar} the phenotypic links of a particular disease are to its links in other layers in the HDMN.
This allows us to derive scores for each disease of how well its phenotypic comorbidities can be explained by genetic, pathway-based, or toxicogenomic mechanisms.
In this sense the derived scores quantify ``how genetic'' or how strong environmental influences are for a given disease.

\section{Data and Methods}

\subsection{Data}

Phenotypic disease-disease associations were obtained from a database of the Main Association of Austrian Social Security Institutions that contains pseudonymized 
claims data of all persons receiving inpatient care in Austria between January 1st, 2006 and December 31st, 2007 \cite{Chmiel14, Thurner13}. 
The data contains age, sex, main- and side-diagnoses (ICD10 codes) \cite{icd} for each hospital stay from $N = 1,862,258$ patients.
Not all ICD codes represent disorders, they may also indicate general examinations, injuries, collections of unspecific symptoms or disorders that are not classified elsewhere.
Unspecific codes are excluded and we work with the remaining $ 1,252$ diagnoses on the three-digit 
ICD levels in chapters (i.e. first-digit-levels) $A$-$Q$, labeled by the capital index $I$.
We use the words disease, disorder and diagnosis interchangeably whenever referring to an ICD entry.

Molecular disease-disease associations were obtained from molecular data of three types, namely purely genetic associations and two different types of environmental associations.
(i) {\it Genetic} disease associations were extracted from the OMIM dataset \cite{OMIM}, which provides a collection of gene-phenotype relationships.
It contains for instance currently more than 30 genes that are known to play a role in type 2 diabetes, e.g. the aforementioned IRS 2 gene.
(ii) {\it Pathway-based} disease associations we took from the UniProtKB database \cite{uniprot, Croft14}.
The UniProtKB database contains protein sequence and functional information that is cross-referenced with pathways in which the proteins play a role and the protein's involvement in diseases.
For instance, an UniProt entry for the PI3-kinase protein cross-references about 40 different pathways, including the PI3K/AKT activation pathway, in addition to three different disease phenotypes from the OMIM dataset.
(iii) {\it Toxicogenomic} disease associations were obtained from the Comparative Toxicogenomic Database (CTD) \cite{Davis14}.
Entries in the CTD correspond to chemicals that are linked to diseases caused by exposure to the substance and with disease genes that are differentially expressed under exposure to it.
For instance, according to this data the chemical methoxychlor is involved in more than ten different diseases, including atopic dermatitis where its influence is mediated by eight different genes, including IL-13.
To link the molecular to the phenotypic data, a mapping between ICD10 and OMIM disease identifiers had to be established.
To obtain such mappings we compiled three different data sources, namely the Human Disease Ontology database \cite{Osborne09}, OrphaNet \cite{Ayme07}, and Wikipedia \footnote{https://en.wikipedia.org/wiki/ICD-10, retrieved 04/30/2015}.
For more information on data extraction and the construction of the ICD10-OMIM mappings see SI, Text S1.
Each of the three molecular datasets can be represented by a bipartite network $B_{i {\textbf j}}^{\alpha}$, where $\alpha$ labels the classes of mechanisms, i.e. genetic ($\alpha = G$), pathway-based ($\alpha = P$), or toxicogenomic ($\alpha = T$), index $i$ labels disorders (ICD10 codes) and {\bf j} labels unique genes (if $\alpha = G$), pathways (if $\alpha = P$), or chemicals (if $\alpha = T$). 
We set $B_{i{\textbf j}}^{\alpha}=1$, if there exists is at least one relation between disease $i$ and gene/pathway/chemical {\bf j}, $B_{i{\textbf j}}^{\alpha}=0$, otherwise.

{\it Heritability and drug approvals.} Information on the broad-sense heritability (see SI, Text S2) of individual diseases $i$, $H_i^2$, was taken from the SNPedia database \cite{SNPedia}.
As a source for drug approvals we used the Drugs@FDA database \footnote{http://www.fda.gov/drugsatfda, retrieved 01/07/2016} from which we obtained FDA-approved brand names and approval dates for all drug products approved since 1939.
These drugs were mapped via known molecular targets to diseases \cite{Yildirim07} to obtain the number of newly approved drug products of the last twenty years for the specific disease $i$, $D_i$.

\subsection{Construction of the HDMN}

We constructed a multi-layer network that encodes disease-disease associations of four different types, the HDMN, $M_{ij}^{\alpha}$.
This network contains one phenotypic layer, $\alpha = \phi$, and three layers that encode molecular disease-disease associations, $\alpha \in \{G,P,T\}$.
The layer of phenotypic disease associations, $M_{ij}^{\phi}$, is given by the contingency coefficient, $\phi_{ij}$, between diseases $i$ and $j$:
Here $N_i$ is the number of patients with disease $i$.
For each pair of diseases $(i,j)$ we counted the number of patients that have both diseases ($N_{ij}$), only disease $i$ or $j$ ($N_{i\bar j}$ or $N_{\bar i j}$, respectively), or neither disease ($N_{\bar i\bar j}$).
Here, the bar denotes ``not''.
Entries in the phenotypic disease network, $M_{ij}^{\phi}$, are then given by the contingency coefficient, 
\begin{equation}
M_{ij}^{\phi} = \phi_{ij} = \frac{N_{ij} N_{\bar i\bar j} - N_{i\bar j} N_{\bar ij}}{\sqrt{N_i N_j (N-N_i) (N-N_j)}} \quad.
\label{ph}
\end{equation}
Values of $\phi_{ij}$ are within the range $\left[-1,+1\right]$ and measure the phenotypic comorbidity strength between diseases $i$ and $j$.
The higher (lower) $\phi_{ij}$, the higher (lower) the probability that a patient with disease $i$ also suffers disease $j$.
$\phi_{ij}=0$ indicates that occurrences of $i$ and $j$ are independent from each other.
We set $M_{ij}^{\phi}=0$, whenever the patient numbers are too low to allow for a reliable estimate of $\phi_{ij}$, i.e. whenever one of the possible outcomes for $N_{ij}$, $N_{i,\bar j}$, $N_{\bar i, j}$, or $N_{\bar i,\bar j})$ was below 5.
An age-dependent version of the phenotypic disease network for a given age interval $t$ is denoted by $M_{ij}^{\phi}(t)$.
Patients fall within one of 11 age groups, 0y-7y, 8y-15y, \dots, 80y-87y.

The layers 2, 3, and 4 of the HDMN encode three different types of molecular associations, $\alpha \in \{G,P,T\}$.
Each of these layers, $M^{\alpha}_{ij}$, is obtained from the bipartite network  $B_{ij}^{\alpha}$ as follows,
\begin{equation}
M_{ij}^{\alpha} = \left\{
    \begin{array}{cl}
      1 & \mathrm{if} \ \sum_{\textbf m} B^{\alpha}_{i{\textbf m}} B^{\alpha}_{j{\textbf m}}>0\\
      
      0 &  \mathrm{if} \ \alpha \in \{P,T\} \ \mathrm{and} \ M_{ij}^G =1\\
      
      0 &  \mathrm{otherwise} 
    \end{array}
  \right. \quad.
  \label{mij}
\end{equation}
Note that this definition ensures that associations between pathologies $i$ and $j$ in the pathway, $M_{ij}^P$, and toxicogenomic, $M_{ij}^T$, layers are indeed due to shared pathways or exposure to the same chemical that can not be explained by direct genetic causes (i.e. $M_{ij}^G =1$).

The numbers of non-isolated nodes, $N^{\alpha}$, and links, $L^{\alpha}$, for each layer $\alpha$ are shown in the SI, table S\ref{data}.
Diseases are not included in the HDMN if they are isolated in every molecular layer $\alpha=G$,$P$, or $T$.
Links in the phenotypic layer $M^{\phi}$ are weighted and typically close to zero \cite{Hidalgo09, Chmiel14}.
Numbers for $N^{\alpha}$ are between 200 and 300 for the molecular layers, whereas there are more than 350 nodes in the phenotypic layer.

\subsection{Disease risks from shared pathophysiological mechanisms}

We introduce a relative risk indicator $RR_i^{\alpha}$ that measures how similar the phenotypic comorbidities of disease $i$ are to its genetic, pathway-based, or toxicogenomic comorbidities.
In this sense $RR_i^{\alpha}$ quantifies how much a specific class of disease-causing mechanisms contributes to the phenotype $i$.
$RR_i^{\alpha}$ is the quotient of the average comorbidity strengths, $M_{ij}^{\phi}$, of all diseases that are linked to $i$ in layer $M_{ij}^{\alpha}$, and the comorbidity strengths of those diseases that are linked to $i$ in none of the pathophysiological layers, i.e.,
\begin{equation}
RR_i^{\alpha} = \frac{\tfrac{1}{k_i^{\alpha}} \sum_j M_{ij}^{\phi} M_{ij}^{\alpha} }{\tfrac{1}{|\mathcal L_i^C|} \sum_j M_{ij}^{\phi} } \quad.
\label{rr}
\end{equation}  
Here $k_i^{\alpha}$ is the degree of disease $i$ in layer $\alpha$ given by $k_i^{\alpha} = \sum_{j} M_{ij}^{\alpha}$ and $\mathcal L_i^C$ is a control set of links for disease $i$ that contains all links $j$, $i \neq j$, for which $\forall \alpha \in \{G,P,T\}: \ M_{ij}^{\alpha}=0 $.
For convenience we also defined the logarithmic relative comorbidity risk, $r_i^{\alpha} = \log RR_i^{\alpha}$.
A value of $r_i^{\alpha}$ close to zero indicates that the presence of pathophysiological comorbidities of type $\alpha$ have no relation whatsoever to the actual, phenotypic comorbidities of $i$.
With increasingly positive values of $r_i^{\alpha}$, the probability increases that the pathophysiological comorbidities of $i$ are indeed observed in the population.

Note that the relative comorbidity risk $r_i^{\alpha}$ can be large due to a single comorbidity $j$ of type $\alpha$ with a very high phenotypic comorbidity strength $M_{ij}^{\phi}$, or because there are a large number of comorbidities with only moderately increased comorbidity strengths.
In particular, $r_i^{\alpha}$ might favor diseases that have a large number of connections of type $\alpha$ to diseases that are physiologically very similar and that have similar ICD10 diagnosis codes, see Text S1.
To adjust for these biases we rescaled $r_i^{\alpha}$ by the node degree $k_i^{\alpha}$ to obtain a measure that favors diseases with a smaller number of highly relevant disease-causing mechanisms.
The re-scaled comorbidity risk, $q_i^{\alpha}$, is given by $q_i^{\alpha} = \tfrac{r_i^{\alpha}}{k_i^{\alpha}}$.

We performed two different statistical tests to evaluate whether $r_i^{\alpha}$ is significantly greater than zero.
First, a Wilcoxon rank sum test for equal medians of two samples was performed.
The samples were given by the set of comorbidity strengths $M_{ij}^{\phi}$ of all diseases $j$ that share a link of type $\alpha$ with $i$, $S_1=\{M_{ij}^{\phi}|\exists j: M_{ij}^{\alpha}=1\}$, and the set $S_2=\{ M_{ij}^{\phi}|j \in \mathcal L_i^C \}$.
The $p$-value for $r_i^{\alpha}$, $p_i^{\alpha}$, was obtained from the one-sided Wilcoxon rank sum test against the alternative hypothesis that the median of $S_1$ is smaller than the median of $S_2$.
A Benjamini-Hochberg multiple hypothesis testing correction was applied on each layer using an exploratory threshold for the false discovery rate of $\alpha = 0.25$ (which corresponds to thresholds for the adjusted $p$-values in the range between 0.1 and 0.05).
Second, we performed a randomization test for $r_i^{\alpha}$ where we replace $M_{ij}^{\alpha}$ by a random permutation of its elements, denoted by $\tilde M_{ij}^{\alpha}$.
The randomized $\tilde{r}_i^{\alpha}$ was computed from equation \ref{rr} where $M_{ij}^{\alpha}$ was replaced by $\tilde M_{ij}^{\alpha}$.
For a given $\alpha$, $\tilde M_{ij}^{\alpha}$ has the same number of nodes and links as $M_{ij}^{\alpha}$, but is otherwise completely randomized.

\section{Results and Discussion}

\begin{figure*}[tb]
\begin{center}
 \includegraphics[width = 0.75\textwidth, keepaspectratio = true]{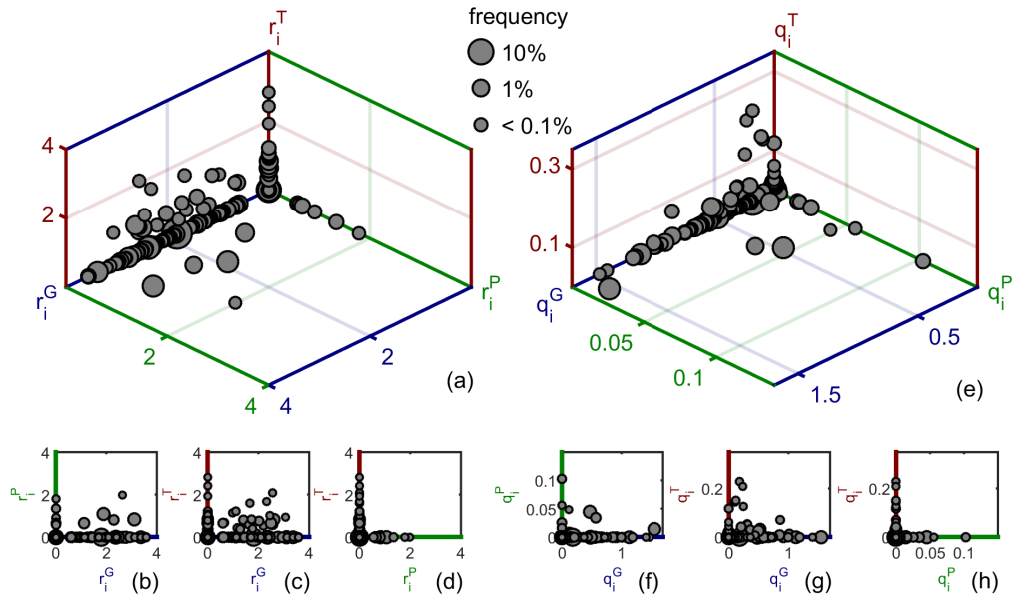}
\end{center}
 \caption{Classification of diseases (circles) according to the dominant causes of their phenotypic comorbidities. Results are shown for (a-d) the relative comorbidity risks $r_i^{\alpha}$ and (e-h) their re-scaled versions, $q_i^{\alpha}$. Circle size is proportional to the number of disease occurrences. Re-scaling the risks by the degrees leads to almost perfect clustering of the diseases around one of the axes. The per-link contribution to the relative comorbidity risk is always dominated by one specific mechanism. Only a comparably small number of diseases cluster around the toxicogenomic axis. The comorbidity risks for most pathologies are dominated by genetic disease-causing mechanisms.}
 \label{fig3}
\end{figure*}

\begin{figure*}[tbp]
\begin{center}
 \includegraphics[width = 0.9\textwidth, keepaspectratio = true]{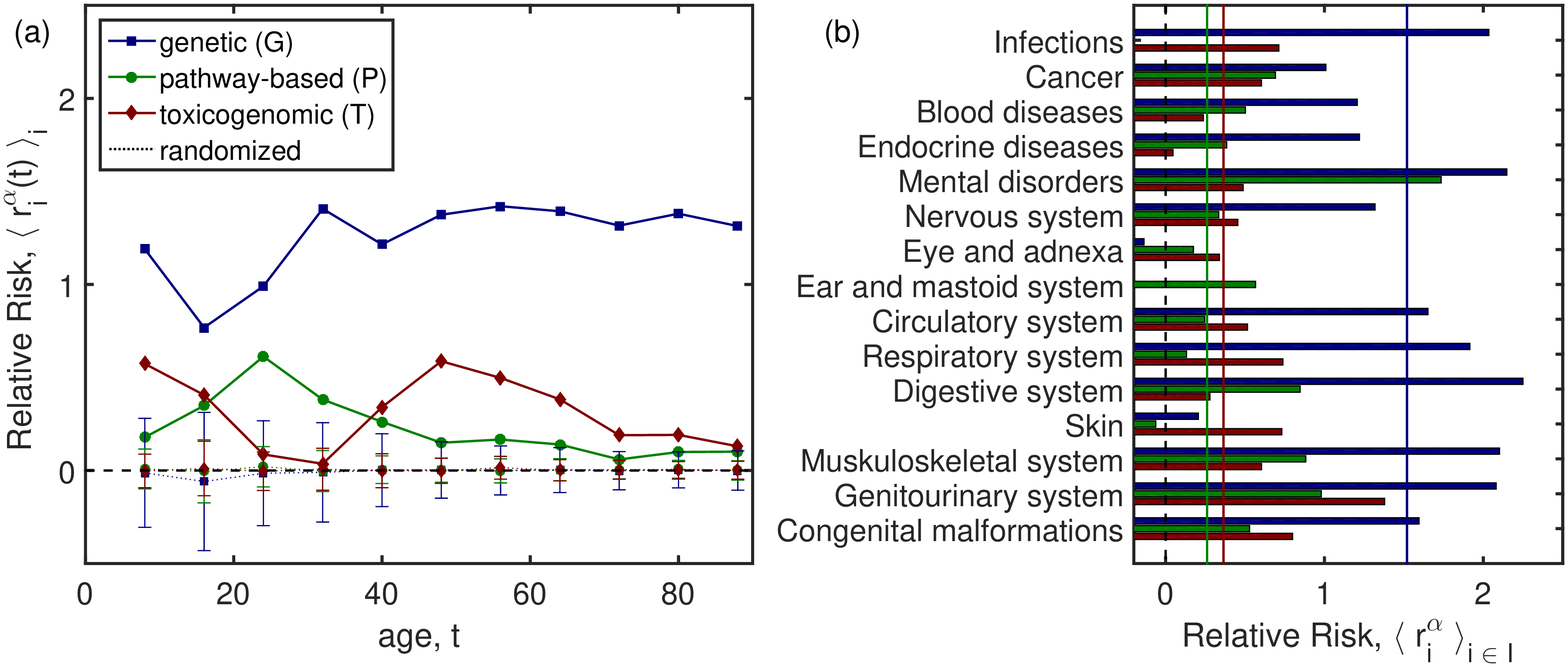}
\end{center}
 \caption{Contributions of genetic, pathway-based, and toxicogenomic comorbidity risks. (a) The genetic risks, $r_i^G(t)$, clearly exceed the pathway-based, $r_i^P(t)$, and toxicogenomic, $r_i^T(t)$, risks across all ages of patients. The results for all three types of mechanisms exceed their expectations from the randomization test (markers connected by dotted lines, error bars show the standard deviation over 5,000 randomizations). (b) Averages of the relative risks are shown for the chapters of the ICD10 classification, the solid vertical lines show the values of genetic (blue), pathway-based (green) and toxicogenomic (red) risks averaged over all diseases. Diseases of the digestive system, mental disorders, and infections show the highest genetically caused comorbidity risk, whereas cancers, diseases of the skin, eye, and ear show the lowest genetic risks. Pathway-based contributions are also highest for mental disorders and toxicogenomic contributions assume their maximum for diseases of the genitourinary system.}
 \label{fig4}
\end{figure*}

The estimates of the most probable disease causes can be visualized in a three-dimensional representation where the axes show the genetic, pathway-based, and toxicogenomic comorbidity risks.
Each disease corresponds to a point with coordinates $(r_i^G, r_i^P, r_i^T)$, see figure \ref{fig3}(a) and its projections onto the (b) $G-P$, (c) $G-T$, and (d) $P-T$ planes.
The size of each marker is proportional to the frequency $N_i/N$ of disease $i$.
We set $r_i^{\alpha}=0$ for all diseases where $r_i^{\alpha}$ is not significantly different from zero after the multiple hypothesis testing correction.
The majority of disorders are clearly dominated by genetic risk factors (many points are close to the $G$-axis).
Some disorders cluster around the $P$ and $T$ axes indicating purely pathway-based and toxicogenomic origins.
Intriguingly, there is precisely no disease that has a significant pathway-based {\it and} toxicogenomic comorbidity risk at the same time, see figure \ref{fig3}(d).
However, a number of disorders with significant pathway-based or toxicogenomic risks have also significant genetic contributions, see figures \ref{fig4}(b) and (c).
This can also be seen in table \ref{top10}, where for instance the chronic nephritic syndrome ranks high in genetic and toxicogenomic comorbidity risks.

The per-link contributions, $q_i^{\alpha}$, of three types of pathophysiological mechanisms are shown in figures \ref{fig3}(e)-(h).
Almost all disorders show one dominant comorbidity risk contribution, i.e. they cluster around a single axis.
Again, most diseases show large genetic risks, while some cluster around the $P$ and $T$ axes.
In the supporting information, SI Figure 2, we show results for $q_i^{\alpha}$ where we allow comorbidities that are at the same time genetic and pathway-based/toxicogenomic (i.e. we drop the second condition for $M_{ij}^{\alpha}$ in equation \ref{mij}).
There are now disorders with, both, significant pathway-based and toxicogenomic comorbidity risks.
For these comorbidities, however, there exists also a direct genetic mechanism that may account for the phenotypic comorbidities.

Table \ref{top10} shows the diseases with the largest genetic, pathway-based, or toxicogenomic comorbidity risks, ranked by statistical significance.
The top genetic diseases include schizo-affective and delusional disorders, as well as schizophrenia.
Different forms of osteoarthritis and chronic bronchitis, as well as nephrotic and nephritic syndromes also show high genetic comorbidity risks. 
The top pathway-based diseases are major depressive disorders, endocrine disorders such as obesity and amyloidosis, diseases of the nervous systems including epilepsy and extrapyramidal and movement disorders, as well as disorders of bone density and multiple myeloma.
The top toxicogenomic diseases include various forms of dermatitis and other skin diseases such as lichen simplex chronicus and prurigo, but also aortic aneurysms, and the chronic nephritic syndrome.

Schizophrenia is indeed a highly heritable disorder that is associated with more than hundred gene loci \cite{Ripke14}. 
The large pathway-based risk for depressions is corroborated by strong and supposedly bi-directional associations between the metabolic syndrome and depression, which have been a long-standing puzzle in epidemiological studies \cite{Pan12}.
Depressions also exhibit strongly significant genetic comorbidity risks ($r_i^G = 2.3$, $p_i^G<0.01$) in consistency with the finding of a gene-by-environment interaction where individuals with a functional polymorphism in the promoter region of the serotonin transporter (5-HT T) gene exhibited more depressive symptoms in relation to stressful life events \cite{Caspi03}.
The high toxicogenomic risks for aortic aneurysms are in line with the effects of chemicals such as nicotine and prostaglandin on related disease-genes \cite{Sakalihasan05}.

\begin{table}[tbp]
\caption{Top 10 diseases in every class of disease-causing mechanisms, $\alpha$, and their relative comorbidity risks $r_i^{\alpha}$, ranked by the significance of its overlap with the phenotypic disease layer, $p_i^{\alpha}$.}
\label{top10}
\begin{tabular}{l p{6cm} l l}
rank  & genetic, $\alpha=G$ & $r_i^{\alpha}$  & $p_i^{\alpha}$ \\
\hline
1	&	F25, Schizo-affective disorders  & 2.4 & $<10^{-4}$	\\					
2	&	F20, Schizophrenia & 2.4 &	$<10^{-4}$	\\		
3	&	M19, Osteoarthritis (unspecified) & 2.9 & $<10^{-4}$	\\		
4	&	N04, Nephrotic syndrome &2.2& $<10^{-4}$	\\		
5	&	J41, Simple, mucopurulent chronic bronchitis &2.1&	$<10^{-3}$	\\					
6	&	J42, Chronic bronchitis (unspecified) &2.0&	$<10^{-3}$	\\							
7	&	M15, Polyosteoarthritis &2.6& 	$<10^{-3}$	\\			
8	&	N03, Chronic nephritic syndrome &2.3&		$<10^{-3}$	\\			
9	&	F22, Delusional disorders &2.6&		$<10^{-3}$	\\			
10	&	M18, Osteoarthritis (first carpometacarpal joint) &2.5&		$<10^{-3}$	\\			
\hline
& pathway-based, $\alpha=P$ &  &  \\
1 &	F32, Major depressive disorder, single episode &1.1&	$<10^{-3}$	\\		
2 &	F33, Major depressive disorder, recurrent &0.81& $0.002$\\
3 &	M85, Disorders of bone density and structure &1.8& $0.003$\\
4 &	G40, Epilepsy and recurrent seizures &0.65&  $0.003$\\
5 &	E66, Overweight and obesity &0.83& $0.006$\\
6 &	E85, Amyloidosis &0.58&	 $0.009$\\
7 &	G25, Other extrapyramidal and movement disorders &0.66&  $0.010$\\	
8 &	H90, Conductive and sensorineural hearing loss &0.56&  $0.010$\\
9 &	M21, Other acquired deformities of limbs &1.3&	 $0.010$\\
10& C90, Multiple myeloma, plasma cell neoplasms &0.90&  $0.011$\\
\hline	
& toxicogenomic, $\alpha=T$ &  &  \\
1 &	I71, Aortic aneurysm and dissection &0.75& $0.002$\\
2 &	L21, Seborrheic dermatitis &0.65& $0.002$ \\
3 &	L24, Irritant contact dermatitis &0.99& $0.002$ \\
4 &	K52, Gastroenteritis and colitis &0.64& $0.002$ \\
5 &	N03, Chronic nephritic syndrome &1.7& $0.004$\\
6 &	L20, Atopic dermatitis &1.2& $0.004$\\
7 &	L28, Lichen simplex chronicus and prurigo &0.69& $0.006$\\
8 &	L30, Unspecified dermatitis &0.58&$0.006$\\
9 &	I89, Noninfective disorders of lymphatic vessels and nodes &0.84& $0.008$\\
10&	G91, Hydrocephalus &0.96 & $0.009$
\end{tabular}
\end{table}

Since phenotypic disease networks are known to undergo large changes in their topology as a function of the age of the underlying patient cohorts \cite{Chmiel14}, we first clarified how the relative comorbidity risks $r_i^{\alpha}$ depend on patient age.
The age-dependent relative risks, $r_i^{\alpha}(t)$, were computed using equation \ref{rr} and by replacing $M_{ij}^{\phi}$ with its age-dependent counterpart, $M_{ij}^{\phi}(t)$.
Results for the average relative comorbidity risks over all diseases $i$, denoted by $\langle r_i^{\alpha}(t) \rangle_i$, are shown in figure \ref{fig4}(a).
Note that this average is also taken over diseases with comorbidity risks $r_i^{\alpha}(t)$ that are not significantly different from zero.
The genetic comorbidity risk averaged over all diseases $i$, $\langle r_i^{G}(t) \rangle_i$, is substantially higher than the pathway-based or toxicogenomic risks and assumes values above 1 for ages between 30 and 90.
Effects are considerably smaller for the average pathway-based (toxicogenomic) comorbidity risks that reach values around 0.5 at ages around 30 (50).
These age differences in the peaks of the environmental comorbidity risks are driven by the age-dependence in the prevalences of the diseases that provide the most dominant contributions to $\langle r_i^{P(T)}(t) \rangle_i$.
In all cases, results for $\langle r_i^{\alpha}(t) \rangle_i$ clearly exceed the expectation values from the randomized risks $\langle \tilde{r}_i^{\alpha}(t) \rangle_i$, obtained from $\tilde M_{ij}^{\alpha}$.
Note that we have confirmed that the dominance of genetic disorders can not be a simple consequence of the exclusion of genetic comorbidities in the other molecular layers in equation \ref{mij}.
Removing this constraint would increase the average environmental contributions by a factor of about 1.5, while the genetic comorbidity risks exceed them by a factor between four and five.
From now on we consider only the time-independent HDMN.

Figure \ref{fig4}(b) shows how much genetic, pathway-based, and toxicogenomic risks contribute to the observed comorbidities for subgroups of diseases that are given by the chapters of the ICD10 classification, the disease groups $I$.
Clear differences between groups of diseases are revealed.
Genetically caused comorbidities include mental disorders, disorders of the digestive system, but also susceptibility to infections.
Genetic mechanisms are least relevant for disorders of the eye, ear, skin, and for cancers.
Pathway-based comorbidity risks are largest for, again, mental disorders and diseases of the genitourinary system.
This shows that the group of mental disorders comprises heterogeneous phenotypes that have either genetically caused or pathway-based comorbidities.
Toxicogenomic comorbidity risks are largest for diseases of the skin, the genitourinary and the respiratory system, as well as for congenital malformations.

\begin{figure}[tbp]
\begin{center}
 \includegraphics[width = 0.45\textwidth, keepaspectratio = true]{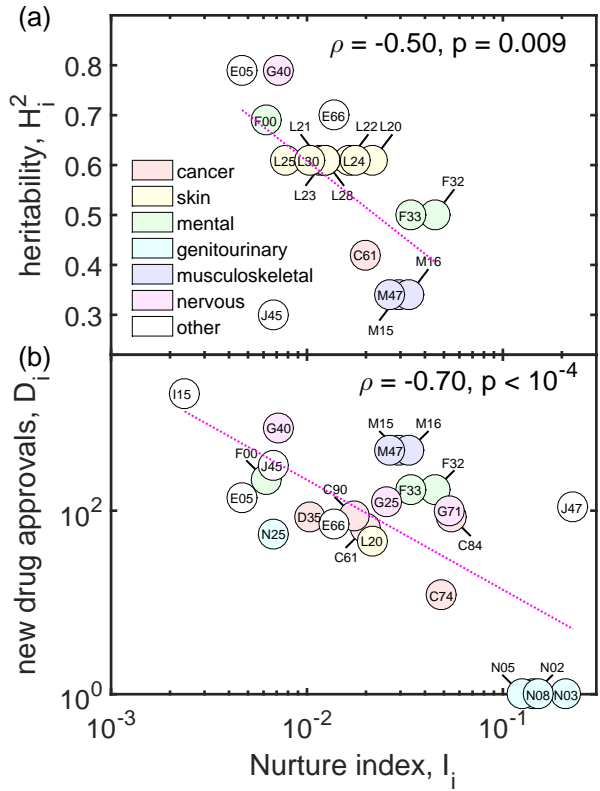}
\end{center}
 \caption{Heritability $H_i^2$, (a) and the number of newly developed drugs $D_i$ (b) are negatively correlated with the relevance of environmental risk factors for diseases. Each circle corresponds to one disease phenotype, labeled by its three-digit ICD10 code. Both, $H_i^2$ and $D_i$ are shown as a function of the nurture index. Colors indicate the main ICD chapter to which the diseases belong. We observe particularly high $I_i$ values for diseases of the genitourinary system, various cancers, depression, and bronchiectasis.}
 \label{fig5}
\end{figure}

The ``nurture index'', $I_i$, quantifies to which extent comorbidities of phenotype $i$ are caused by environmental, i.e. pathway-based or toxicogenomic, mechanisms,
\begin{equation}
I_i = \sqrt{(q_i^P)^2 + (q_i^T)^2} \quad.
\label{ngi}
\end{equation}
Figure \ref{fig5} shows results for (a) the heritability and (b) the number of new drug approvals $D_i$ as a function of $I_i$.
Each circle in figure \ref{fig5} corresponds to a disease phenotype, labeled by its ICD10 code.
The colors of the circles refer to their chapter in the ICD classification. 
The highest values of $I_i$ are found for diseases of the genitourinary system (N03 and N05 nephritic syndrome, N02 hematuria, N08 glomerular disorders), depressions (F32, F33), several cancers (C84 T/NK-cell lymphoma, C74 adrenal gland, C61 prostate), as well as bronchiectasis (J47).
Figure \ref{fig5}(a) shows that there is a significant negative correlation between the nurture index, $I_i$, and the broad-sense heritability, $H_i^2$, of disorder $i$.
This corroborates that $I_i$ is indeed related to the plasticity of phenotype $i$, i.e. $I_i$ increases with the influence of environmental risk factors.
There is also a strong significant negative correlation between the logarithms of $I_i$ and $D_i$ shown in figure \ref{fig5}(b).
We found this result to be very robust for a large variety of choices of this time span, ranging from five years upwards.
Note that $\log D_i$ and $H_i^2$ show no significant correlation among them ($\rho = 0.19$, $p=0.17$).
This indicates a significant bias in pharmaceutical R\&D that favors market placements of drugs that target disorders with low environmental risk factors.
It has indeed been shown that the success rates for drug development vary dramatically among disease areas \cite{Nelson15}.
These rates have been found to increase with the existence of direct genetic evidence, which in particular applies to diseases of the musculoskeletal system and infections, which we also identified as predominantly genetic in figure \ref{fig3}(b).

\section{Conclusions}

We developed a novel approach to quantitatively disentangle the most relevant genetic or environmental disease-causing mechanisms for a large number of particular disorders.
This has become possible through recent advances in observing networks of phenotypic comorbidity relations with unprecedented precision \cite{Hidalgo09, Chmiel14}.
We considered three different classes of mechanisms that can be at the core of these observed comorbidities, namely genetic, pathway-based, and toxicogenomic mechanisms that cause more than one disorder.
By constructing the HDMN we have been able to identify the most probable causes for 358 different phenotypes by measuring the overlap between phenotypic and pathophysiological comorbidities, the relative comorbidity risks $r_i^{\alpha}$.
We find that the different environmental disease-causing mechanisms do not mix; we found no pathologies that have significant pathway-based {\it and} toxicogenomic comorbidity risk contributions at the same time.
While for most of the studied diseases genetic risk factors dominate, we identify a number of disorders with significant environmental contributions which typically coincides with low heritability and lower rates of successful market placements of drugs.

Our approach cross-validates pathophysiological mechanisms by whether their predicted comorbidities are indeed directly observed in the population.
Moreover we can rule out certain types of disease-causing mechanisms when the comorbidities that they predict are not observed.
The methodology developed here can be extended to decide on a quantitative basis if the comorbidities predicted by a particular {\it individual} pathophysiological mechanism are also phenotypically relevant.
The new technology can be used as a novel and data-driven way to validate potential drug targets.

\subsection{Acknowledgments}
We are very grateful to J\"org Menche for stimulating discussions and acknowledge financial support from the European Commission, FP7 project MULTIPLEX No. 317532.

\pagebreak

\section{Supplementary Information}
\setcounter{table}{0}
\setcounter{figure}{0}
\subsection{Text S1: Further details on data extraction and MIM-ICD10 mappings}

Genetic disease associations are extracted from the Online Mendelian Inheritance in Man (OMIM) dataset, from which we obtained a list of 4,847 associations between disorders (phenotype MIM numbers) and genes \cite{OMIM}.
Thereby we included only those phenotype-gene associations for which the molecular basis of the disorder is known (i.e. a phenotype mapping key with value 3 in the OMIM dataset).
{\it Metabolic} disease associations stem from the UniProtKB database, which provides a list of 3,020 proteins that are known to be involved in disorders in humans (given by phenotype MIM numbers) \cite{uniprot}.
The REACTOME database cross-references these proteins with pathways in which they occur \cite{Croft14}.
{Toxicogenomic} disease associations are obtained from the Comparative Toxicogenomic Database as a list of 4,925 associations between disorders (phenotype MIM numbers or MeSH ID) and chemicals \cite{Davis14}.
Here we only use curated disease-chemical associations, i.e. those for which direct, literature-curated evidence exists.

To obtain mappings from MIM phenotype numbers and MeSH codes to the ICD10 classification we compile three different data sources.
In addition to the Human Disease Ontology database \cite{Osborne09} and mappings provided from OrphaNet \cite{Ayme07}, we extracted mappings by crawling a disease index page from Wikipedia (https://en.wikipedia.org/wiki/ICD-10, retrieved 04/30/2015).
From these three sources result 85,303 MeSH-ICD10 and 5,498 MIM-ICD10 associations.
Aggregated to the three-digit ICD10 level, $90\%$ of the ICD10 codes can be mapped to MeSH codes and $37\%$ to MIM numbers.
This lower number of successfully translated MIM numbers is partly due to the fact that not for all diseases a molecular basis is known or even relevant.

While the ICD10 codes are primarily used for billing and clinical purposes, the OMIM classification focuses on descriptive phenotypes of inherited conditions.
From this follows the limitation that some very specific OMIM codes might link to highly unspecific ICD10 codes and {\it vice versa}.
For instance, colorectal cancer has one MIM number (114500) but four different ICD10 codes on the three-digit level, C18-C21.
These four phenotypes are not only connected among each other; each disease that is genetically linked to colorectal cancer is also linked to all four of these ICD10 codes.
Consequently the diagnoses C18-C21 have the highest degrees in the genetic comorbidity network.
Similarly, the ICD10 codes for essential hypertension (I10-I13) all map to a single MIM number and have the highest degrees in the toxicogenomic disease network.
We therefore adjusted for such biases by re-scaling the relative comorbidity risk $r_i^{\alpha}$ by the node degree $k_i^{\alpha}$.

\subsection{Text S2: Broad-sense heritability}
Heritability is a measure that quantifies how much variation in a phenotypic trait (such as a disease) in a population is due to genetic variation among individuals in the population. More specifically, if $\sigma_G$ is the genetic variation and $\sigma_P$ the variation in the population, the broad-sense heritability, $H^2$, is defined as $H^2 = \tfrac{\sigma_G}{\sigma_P}$. Sloppily defined, heritability measures the proportion of (disease) risk that is due to the genetic background of an individual, as opposed to environmental factors. However, high values of heritability do not necessarily imply a high disease risk, as it may be relatively easy to prevent certain genetic diseases by certain interventions.

\begin{table}[tbp]
\caption{Overview of characteristics of the HDMN layers. The numbers of non-isolated nodes, $N^{\alpha}$, and links, $L^{\alpha}$, are given for four different layers.}
\label{data}
\begin{tabular}{l l l}
$\alpha$ & $N^{\alpha}$ & $L^{\alpha}$ \\
\hline
phenotypic & $358$ & $63903$\\
genetic & $285$ & $969$\\
pathway-based & $251$ & $3930$\\
toxicogenomic & $199$ & $4994$\\
\end{tabular}
\end{table}

\begin{figure}[tb]
\begin{center}
 \includegraphics[width = 0.45\textwidth, keepaspectratio = true]{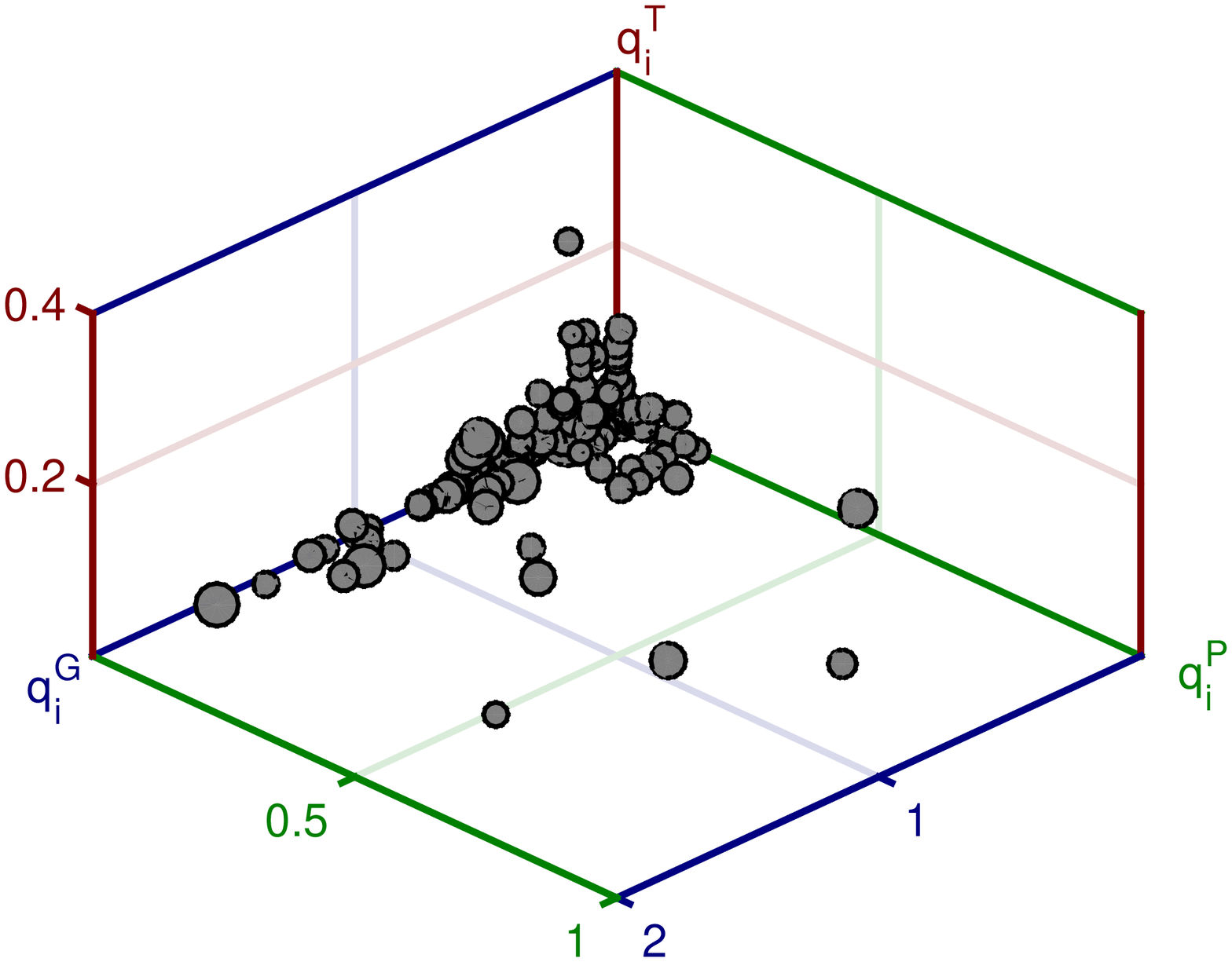}
\end{center}
 \caption{Classification of diseases (circles) according to the dominant contributions to their phenotypic comorbidities for an alternative definition of the re-scaled relative comorbidity risks, $q_i^{\alpha}$, where two diseases can at the same be comorbid in a genetic of pathway-based / toxicogenomic way. Again, most diseases cluster around one of the axis with a clear dominance of genetic comorbidity risks.}
 \label{SIfig1}
\end{figure} 

\end{document}